\documentclass[prl,longbibliography]{revtex4-1}
\usepackage{amsmath}
\usepackage{amssymb}
\usepackage{graphicx}
\usepackage{epsfig}
\usepackage{bm}
\usepackage{color}
\newfont{\sss}{cmmi10 at 20.74pt}

\def\spose#1{\hbox to 0pt{#1\hss}}
\def\simlt{\mathrel{\spose{\lower 3pt\hbox{$\mathchar"218$}}
     \raise 2.0pt\hbox{$\mathchar"13C$}}}
\def\simgt{\mathrel{\spose{\lower 3pt\hbox{$\mathchar"218$}}
     \raise 2.0pt\hbox{$\mathchar"13E$}}}
\def\0{\mbox{\boldmath$\displaystyle\mathbf{0}$}}

\def\be{\begin{equation}}
\def\ee{\end{equation}}
\def\bea{\begin{eqnarray}}
\def\eea{\end{eqnarray}}

\newcommand{\units}[1]{\ensuremath{\,\mathrm{#1}}}
\font\sc=cmcsc10

\begin{document}

%\textbf{\large  Quantum entanglement induced apparent CP violation in neutrino oscillations }
\textbf{\large Muon lifetime dependent effects in MiniBooNE and LSND}

\vspace{21pt}

{\sc D. V. Ahluwalia, S. P. Horvath}\\ 
\textit{\small Department of Physics and Astronomy\\
Rutherford Building, University of Canterbury \\ 
Private Bag 4800, Christchurch 8020, New Zealand}
\\ \\
\textit{E-mail: dharamvir.ahluwalia@canterbury.ac.nz}

\vspace{31pt}
 \begin{abstract}
\vspace{15pt}
 We argue that because  the source-detector distance of $1.8 \units{\mbox{$\mu$s}} $ (in natural units) for the MiniBooNE is comparable to the   muon lifetime of $2.2 \units{\mbox{$\mu$s}}$, and because time dilation effects are wiped out in the beam stop, the Goldman entanglement of the neutrinos  leads to hitherto unsuspected interpretational consequences. We show that a distinct possibility exists in which a LSND-like experiment  sees no CP violation, whereas a MiniBooNE-like setup reproduces the LSND results for the $\bar\nu_\mu$ to $\bar\nu_e$ oscillations while seeing only a significantly suppressed signal for the $\nu_\mu$ to $\nu_e$ oscillations. We also discuss an alternate scenario. This also suggests that the LSND  experiment and the MiniBooNE  should not be compared without taking into account the Goldman entanglement.

\end{abstract}

\maketitle
\textit{Introduction \textemdash}
Whenever a mass eigenstate decays into a set of other mass eigenstates the relevant conservation laws may induce a quantum entanglement. For instance, in the EPR-like decay of a scalar into two spin one half particles (in the singlet state), the conservation of angular momentum forces the spin projections of the decay products to be entangled. Such an entanglement is quite robust~\cite{PhysRevA.76.052110} and it is destroyed, e.g., when the entangled attribute is subjected to a measurement or is otherwise destroyed. Another type of quantum entanglement may occur if  one of the decay products is a superposition of different mass eigenstates. Since flavour eigenstates of neutrinos are a superposition of different mass eigenstates,  the energy-momentum conservation induces a variety of quantum entanglements in the  decay products accompanying such neutrinos~\cite{Goldman:1996yq,Nauenberg:1998vy,Cohen:2008qb}. The backaction induced by the propagation of the decay products through the beam stop of a neutrino oscillation experiment causes the decoherence of the mass-eigenstate superpositions.  If, during the time it takes a neutrino to traverse the source-detector distance, the entangled  particles  decay or are annihilated, then the backaction could be considered strong with an expectation of a concrete signature in the neutrino detector. Since the source-detector distance of $1.8 \units{\mbox{$\mu$s}} $ (in natural units) for the MiniBooNE is comparable to the muon lifetime of $2.2 \units{\mbox{$\mu$s}}$, the indicated entanglement  must be incorporated in any interpretation of this  experiment (note: it takes only a few nanoseconds for the muons to come to rest in the beam stop). An important contribution to the backaction under discussion may reside in the  fact that the decay products of the entangled particles encounter a CP asymmetric beam stop. This opens up an annihilation channel for  positrons (produced, e.g., in the decay of an entangled $\mu^+$) and this serves as a strong candidate for the indicated decoherence of the mass-eigenstate superpositions. Consequently this circumstance may be mistakenly interpreted as an intrinsic CP violation. In contrast, for the LSND experiment the source-detector distance is only about $0.1 \units{\mbox{$\mu$s}} $. For this reason, the entanglement, and apparent CP violation, effects become negligible for the LSND experiment.\\

\textit{Quantum entanglement and the role of muon lifetime \textemdash}
In 1996 the LSND experiment at the Los Alamos Meson Physics Facility reported evidence for $\bar{\nu}_\mu \rightarrow \bar{\nu}_e$ oscillations~\cite{Athanassopoulos:1996jb}. For the used setup, the observed oscillation probability was $\mathcal{P}^{\text{LSND}}_{\bar{\nu}_\mu\to \bar{\nu}_e} = (0.31 \pm 0.12 \pm 0.05)\times 10^{-2}$. Two years later~\cite{Athanassopoulos:1997pv}, the LSND collaboration provided a lower statistics result for the  ${\nu}_\mu \rightarrow {\nu}_e$ oscillations with $\mathcal{P}^{\text{LSND}}_{{\nu}_\mu\to {\nu}_e} = (0.26 \pm 0.10 \pm 0.05)\times 10^{-2}$. Since these results were in conflict with the part of the parameter space explored by the  KARMEN experiment and compatible for the remaining parameter space (especially for $\Delta m^2 \le 2 \units{\mbox{eV$^2$}} $~\cite{Eitel:1999gt}), the MiniBooNE collaboration was formed to resolve the issue. The first results from MiniBooNE are inconclusive on $\bar{\nu}_\mu \rightarrow \bar{\nu}_e$ oscillations, anomalous with respect to ${{\nu}_\mu\to {\nu}_e}$, and null at the $90\%$ level for the $\nu_\mu$ and $\bar{\nu}_\mu$ disappearance~\cite{AguilarArevalo:2009xn,AguilarArevalo:2008rc,AguilarArevalo:2009yj}. \\

To initiate an examination of these perplexing experimental results, we will take it as a working hypothesis that none of the experiments suffer from a serious flaw. Instead, we will take the view that the interpretation of the experiments is missing an unsuspected piece of known physics. This view is justified in view of the readily identifiable differences between these two experiments. These differences are related to the above-noted pion lifetime and shall become apparent below.
 \\

 To identify these missing physics elements and to explore their implications, consider the production of $\nu_\mu$ and $\bar{\nu}_\mu$  in the following CP conjugated processes 
\begin{equation}
  \begin{picture}(70,30)(100,-20)
    \put(0,0){$\pi^+$}
    \put(15,2.5){\vector(1,0){15}} 
    \put(35,0){$\nu_\mu + \mu^+$}
    \put(60,-7.5){\line(0,-1){10}}
    \put(60,-17.5){\vector(1,0){10}}
    \put(80,-20){$ e^{+} + \nu_\text{e} + \bar{\nu}_\mu \;,$}
  \end{picture}
  \begin{picture}(70,30)(0,-20) % 140,30 picture size in pt; -20,-20 moves center left 20 pt
    \put(0,0){$\pi^-$} %origin after moving
    \put(15,2.5){\vector(1,0){15}} % 15 start of arrow, 2.5 takes it 2.5 pt up \put{where}{what} (1,0)= direction horizontal right, {-1,0}=horizontal left,{1,1} = up etc. Last enter 15 is 15 pt length of the vector
    \put(35,0){$\bar{\nu}_\mu + \mu^-$}
    \put(60,-7.5){\line(0,-1){10}}
    \put(60,-17.5){\vector(1,0){10}}
    \put(80,-20){$ e^{-} + \bar{\nu}_\text{e} + {\nu}_\mu $}
  \end{picture}\label{eq:piondecay}
\end{equation}
The neutrinos and antineutrinos that are thus produced are a linear superposition of different mass eigenstates. For this reason, as first argued by Goldman~\cite{Goldman:1996yq} (and later emphasised by Nauenberg on the one hand~\cite{Nauenberg:1998vy}  and Cohen, Glashow,  and Ligeti on the other hand~\cite{Cohen:2008qb}), the conservation of the energy-momentum four vector induces a quantum entanglement between the decay products. For the  $\pi^\pm$ decay at rest, the neutrino-muon entanglement becomes manifest by re-writing $\nu_\mu + \mu^+$ and $\bar{\nu}_\mu + \mu^-$ in Eq.~(\ref{eq:piondecay}) as
\begin{equation}
  \sum_i U_{\mu i} \underbrace{\left\vert \nu_i\right\rangle}_{\left\vert\sqrt{\mathbf{k}_i^2+m_i^2},\;\mathbf{k}_i,\;m_i^2\right\rangle} \otimes \underbrace{\left\vert\mu^+\right\rangle}_{\left\vert E_i,\;-\mathbf{k}_i,\; m_\mu^2\right\rangle} ,\qquad \sum_i U^\ast_{\mu i} \underbrace{\left\vert \bar{\nu}_i\right\rangle}_{\left\vert \sqrt{\mathbf{k}_i^2+m_i^2},\;\mathbf{k}_i,\;m_i^2\right\rangle} \otimes \underbrace{\left\vert\mu^-\right\rangle}_{\left\vert E_i,\;-\mathbf{k}_i,\; m_\mu^2\right\rangle}\label{eq:entanglement}
\end{equation}
The notational details are adapted from Ref.~\cite{Goldman:1996yq}. Once the Goldman suggestion is taken seriously one must in fact also incorporate the entanglement induced by the conservation of angular momentum. This forces the above states to be replaced by their proper singlet state expressions which contain not only the left-transforming neutrino with the negative helicity, but also the left-transforming neutrino  with the positive helicity. The latter may mimic some of the signatures of a sterile neutrino (similar remarks hold true for antineutrinos). This is a subject that must be treated in a subsequent study. In order not to raise too many issues at once the rest of the paper takes Eq.~(\ref{eq:entanglement}) without any corrections.\\

As long as the decay products have an infinite lifetime and are not subjected to an interaction with the environment the Goldman analysis guarantees the standard expression for the flavour-oscillation probabilities (to the lowest order in the underlying mass-squared differences). Our point of departure shall  reside in the observation that the evolution of the neutrino flavour states is intricately tied to the environment that the entangled muons encounter and whether the neutrinos reach the detector prior to the decay of the associated muons. For concreteness, we now enumerate the elements of physics that distinguish the two experiments under examination
\begin{enumerate}
\item  Let $\ell$ and  $\ell^\prime$ represent the source-detector distances for the MiniBooNE and the LSND experiments. In natural units (as already noted), $\ell= 1.8 \units{\mbox{$\mu$s}} \approx \tau_\mu$, whereas  $\ell^\prime = 0.1 \units{\mbox{$\mu$s}}\approx 0.05 \tau_\mu $;  where $\tau_\mu \approx 2.20  \units{\mbox{$\mu$s}}$ is the mean lifetime of $\mu^\pm$. Consequently, for the LSND experiment most of the  entangled muons decay after the associated neutrinos pass through the detector. For the experiment of the MiniBooNE collaboration, the opposite holds; that is, a significant number of muons decay before the associated neutrinos arrive at the detector.  
\item The propagation of the $\mu^+$ as well as the $\mu^-$ through the beam stop affects the entangled $\nu_\mu$ and $\bar{\nu}_\mu$.  When the former decays it produces a $e^+$, while the decay of the latter gives an $e^-$. These $e^\pm$ encounter a CP asymmetric beam stop thus opening up an annihilation channel for the $e^+$, but not for the $e^-$. The decoherence transfer to the environment and loss of coherence for  $\nu_\mu$ and $\bar{\nu}_\mu$ thus becomes asymmetric. The consequences of this circumstance may be mistakenly interpreted as an intrinsic CP violation.
\end{enumerate}
With these observations in mind it is clear that the LSND neutrino oscillation results are, by and large, free from the effects induced by Goldman entanglement of $\nu_\mu$ ( $\bar{\nu}_\mu$) with $\mu^+$ ($\mu^-$). The LSND has been dismantled but its last results showed no signs of CP violation
\begin{equation}
\mbox{LSND:}\quad\mathcal{P}(\bar{\nu}_\mu\to\bar{\nu}_e) = \mathcal{P}({\nu}_\mu\to{\nu}_e)
\end{equation}
On the other hand MiniBooNE is still running and new results are  expected soon. Given the above observations and taking hints from the Martin\textendash Zurek analysis~\cite{PhysRevLett.98.120401} which studied transfer of entanglement to the environment and argued that backaction of the environment causes decoherence of energy-eigenstates superpositions, we conjecture (see below) the following phenomenological modification to the canonical neutrino oscillation probability
\begin{align}
 &\chi\left(\bar{\nu}_\mu\to\bar{\nu}_e\right) =
 \mathcal{P}\left(\bar\nu_\mu\to\bar\nu_e\right)\label{eq:zimpokb}
 \\
 & \chi\left(\nu_\mu\to\nu_e\right) = \exp[-\ell/\tau_\mu] 
 \mathcal{P}\left(\nu_\mu\to\nu_e\right) +  \left(1- \exp[-\ell/\tau_\mu]\right) \sum_i \left(U^\ast_{\mu i} U_{\mu i} U_{e i} U^\ast_{ei} \right)\label{eq:zimpoka}
\end{align}
The above modification, to be dubbed `hard',  takes the view that the annihilation of $e^+$  \textendash~see the decay chain of $\pi^+$ in (\ref{eq:piondecay})  \textendash~has the effect of destroying the entanglement of $\nu_\mu$ and $\mu^+$~\footnote{Without this annihilation,  we envision, the entanglement could simply propagate like a fog of energy-conserving processes throughout the beam stop. For the decay of $\mu^-$ in the $\pi^-$ chain we assume that the entangled $\mu^-$ transfers its entanglement properties to the $ e^{-} + \bar{\nu}_\text{e} + {\nu}_\mu $. Any loss of entanglement is slower, and we  dub it as `softer.' We do not explicate the `softer' effects any further}. If a Martin and Zurek like argument holds, this backaction decoheres the flavour states to the underlying mass eigenstates. The $ \exp[-\ell/\tau_\mu] $ factor controls the number of $e^+$ that undergo annihilation in the beam dump. In Eq.~(\ref{eq:zimpoka}) $U^\ast_{\mu i} U_{\mu i} U_{e i} U^\ast_{ei} $ contains the probability of projecting the $i$th mass eigenstate in the backaction caused decoherence of $\nu_\mu$ multiplied by the probability that the flavour detector detects it as a $\nu_e$. \\

In the `hard'  version of our conjecture LSND sees no CP violation, while MiniBooNE reproduces the LSND results for the $\bar\nu_\mu$ to
$\bar\nu_e$ oscillations while seeing only a significantly suppressed signal for the $\nu_\mu$ to $\nu_e$ oscillations. The oscillatory term
now has roughly $44\%$ of the expected amplitude while the signal develops a roughly $56\%$ constant addition to the $e$-like flux. The actual number of events depends on the energy-dependent neutrino cross section and the detector efficiency. Should MiniBooNE provide an indication for such a scenario then the existing neutrino oscillation data and its analysis would require a significant new effort to settle the mass-squared differences and the mixing matrix~\cite{Vissani:1997pa,Ahluwalia:1998xb,Barger:1998ta,Altarelli:1998sr,Baltz:1998ey,Stancu:1999ct,Harrison:2002er,Xing:2002sw,Georgi:1998bf,Xing:2001cx}. We have confirmed this suspicion by analysing such a hypothetical MiniBooNE scenario in conjunction with data from SN1987a~\cite{PhysRevD.65.063002}. The reason for making this choice of the data is simple. Each set of data comes with its set own quantum entanglements. These are far from simple to model. For instance, the entanglement and its evolution for the atmospheric neutrinos differs from bin to bin. The bin containing the data for the zenith angle at zero suffers from no backaction decoherence whereas the opposite is true for the bin with the zenith angle at around $\pi$ (there are also energy-dependent effects arising from the time dilation of the muon lifetime). For reactor neutrinos, one must account for the entangled neutron inside the heavy nuclei. The dominant source of $\bar{\nu}_e$ production are the following nuclear reactions and decays
\begin{equation}
  \begin{picture}(140,50)(80,-40)
    \put(-75,0){ $n + {}^{238}\text{U} $}
    \put(-30,2.5){\vector(1,0){15}} 
    \put(-12.5,0){${}^{239}\text{U} \text{ (lifetime 23.5 min)}$}
    \put(95,2.5){\vector(1,0){15}}
    \put(115,0){$ {}^{239}\text{Np} + e^{-} + \bar{\nu}_e $  } 
    \put(135,-7.5){\line(0,-1){10}}
    \put(135,-17.5){\vector(1,0){10}}
    \put(155,-20){${}^{239}\text{Pu} +e^- + \bar{\nu}_{e}$ \text{(lifetime 2.36 days)}}
    \put(175,-27.5){\line(0,-1){10}}
    \put(175,-37.5){\vector(1,0){10}}
    \put(190,-40){${}^{235}\text{U} + {}^4\text{He}$ \text{(lifetime $2.4 \times10^4$ years)}}
  \end{picture}
  \label{eq:nr1}
\end{equation}
Nuclear densities inside uranium and neptunium are similar to those of neutron stars. If the time scale of the backaction induced decoherence becomes comparable to the transit time of neutrinos from the reactor to the detector then the standard expectations would need a significant revision. To compound these new elements of largely unexplored physics one must also merge MSW matter effects. For these reasons we have chosen to narrow our focus to the cases which seemed easiest to analyse.\\

\textit{Analysis of LSND, MiniBooNE, and SN1987a~\textemdash}
For the SN1987a data, we used the events for Kamiokande II and IMB as summarised in Ref.~\cite{PhysRevD.65.063002}. Taking into account the distance of the SN1987a, we concluded that each of the flavour eigenstates spreads out spatially. So, the flavour detectors project a mass eigenstate to a flavour eigenstate. Thus the energy of an event tags whether the event arises from the mass eigenstate associated with an $e$-like neutrino, or from one of the other flavours. Type II Supernova models typically suggest that  $\nu_e$ and $\bar\nu_e$ energies $\lesssim 15\units{MeV}$, while the other two flavours have energies above this value. We exploited this fact to bin the events~\textendash~all of which are $e$-like~\textendash~into two bins: one collecting the events that are likely to have been projected from the $e$-like flavour (and carry an  energy of $\lesssim 15\units{MeV}$) and the other where the events are likely to have arisen from the $\mu$- and $\tau$- like neutrinos (and carry an energy of $\gtrsim 15\units{MeV}$). We term the former as the ``$e$-bin,'' and the latter as ``$\mu\tau$-bin.'' The theoretically expected events in each bin are summarised in table~\ref{tab:zimpok}.\\ 

\begin{table}[hbt!]
  \centering
  \begin{tabular}{|c || c  c c|}
    \hline
    Neutrino Flavour & $e$-bin & $\mu\tau$-bin & Data exists \\
    \hline\hline
    $\nu_e$ and $\bar{\nu}_e$ flux & \textcolor{black}{$N^{\text{SN}}_e \sum_i U^\ast_{ei} U_{ei} U_{ei} U^\ast_{ei}$ }&  \textcolor{black}{$N^{\text{SN}}_\mu \sum_i U^\ast_{\mu i} U_{\mu i} U_{ei} U^\ast_{ei}$} +  \textcolor{black}{$N^{\text{SN}}_\tau \sum_i U^\ast_{\tau i} U_{\tau i} U_{ei} U^\ast_{ei}$} & Yes\\ 
    $\nu_\mu$ and $\bar{\nu}_\mu$ flux & $N^{\text{SN}}_e \sum_i U^\ast_{ei} U_{ei} U_{\mu i} U^\ast_{\mu i}$ &  $N^{\text{SN}}_\mu \sum_i U^\ast_{\mu i} U_{\mu i} U_{\mu i} U^\ast_{\mu i}$ +  $N^{\text{SN}}_\tau \sum_i U^\ast_{\tau i} U_{\tau i} U_{\mu i} U^\ast_{\mu i}$ & Not yet\\ 
    $\nu_\tau$ and $\bar{\nu}_\tau$ flux &  $N^{\text{SN}}_e \sum_i U^\ast_{ei} U_{ei} U_{\tau i} U^\ast_{\tau i}$ &  $N^{\text{SN}}_\mu \sum_i U^\ast_{\mu i} U_{\mu i} U_{\tau i} U^\ast_{\tau i}$ +  $N^{\text{SN}}_\tau \sum_i U^\ast_{\tau i} U_{\tau i} U_{\tau i} U^\ast_{\tau i}$ & Not yet\\
    \hline
  \end{tabular}
  \caption{$N^{SN}_f = n^{SN}_f \times g$ (see definitions after Eq.~(\ref{eq:zimpokw})), where $g$ is a geometrical factor and $f$ represents the flavour. The number of events is then calculated by appropriately taking into account the variation of neutrino cross sections with energy and including the detector efficiencies. In our calculation we used the detector efficiency of unity and the ratio of cross sections for the $e$-bin events was approximated as (11/31) times that of the $\mu\tau$-bin.  The geometrical  factor $g$ differs from detector to detector. Here we treat the data from IMB and Kamiokande II as a single combined data set with the same $g$.
 Further details appear in the text.} 
 \label{tab:zimpok}
\end{table} 
\noindent The average energy for  events in the $e$-bin was determined to be $9\units{MeV}$ (as carried by the positron in the detector). This roughly translates to $11\units{MeV}$ for the average neutrino energy~\cite[p. 420]{Raffelt:1996wa}. Similarly, the average neutrino energy associated with the events in the $\mu\tau$ bin was obtained to be $31\units{MeV}$. We then assumed that for SN1987a the energy flux was equally divided between the three flavours. This gave us 
\begin{equation}
  n^{SN}_{\mu\tau} \approx \frac{11}{31} n^{SN}_e\label{eq:zimpokw}
\end{equation}
where $n^{SN}_{\mu\tau}$ is the combined number of $\mu$ and $\tau$ type neutrinos at the source, and $n^{SN}_e$ represents the $e$-like counterpart. On using table \ref{tab:zimpok}, these observations  yield the \textit{theoretical} expression for the ratio of the number of events in the $\mu\tau$-bin to the 
 number of events in the $e$-bin 
 \begin{equation}
   \gamma^{th}\approx  \frac{31 \big[9 -7 \cos(2 \theta_{13}) -  2 \cos(4\theta_{12}) \cos^2(\theta_{13}) \big]  \cos^2(\theta_{13})}{22 \big[ 4 \sin^4(\theta_{13}) + (3 + \cos(4 \theta_{12}) )\cos^4(\theta_{13}) \big]}
 \end{equation}
In obtaining the above expression we assumed that the neutrino cross section in the detector was proportional to $E_\nu^2$; therefore, the events in the $e$-bin the neutrino flux was multiplied by $(11)^2$, while for the $\mu\tau$-bin the multiplicative factor was $(31)^2$.
 Equating  $\gamma^{th}$ to its \textit{observed} value, $\gamma$, determines $\theta_{13}$ in terms of $\theta_{12}$
\begin{equation}
  \theta_{13} \approx \cos^{-1}\left[ \frac{22 \gamma}{(62+ 22 \gamma) - \sqrt{(31+11\gamma)(124 -33\gamma- 11 \gamma\cos(4 \theta_{12}))  }}   \right]^{1/2}
\end{equation}
Binning the data in the indicated manner, we found that the $e$-bin as well as the $\mu\tau$-bin contains 12 events each; giving  $\gamma =1$. This simplifies the above expression to
\begin{equation}
  \theta_{13} \approx \cos^{-1}\left[\frac{22}{84-\sqrt{42}\sqrt{91 - 11 \cos(4 \theta_{12})} }\right]^{1/2}\label{eq:theta13}
\end{equation}
Having thus obtained a parametric constraint on $\theta_{12}$ and $\theta_{13}$ we proceeded to 
find a common parameter space for the  MiniBooNE and the LSND experiment. We assumed that the mass-squared difference associated with these experiments was  $\gg$  the sole mass-squared difference that is left in a $3\times 3$ analysis with no intrinsic CP violation. That is, in the absence of a sterile neutrino, we
expect that when a full analysis of the neutrino oscillation data is done, along the lines we have outlined, both the solar and atmospheric data will require only one mass-squared difference. With $\ell$ set to $541\units{m}$, we used Eq.~(\ref{eq:zimpokb}) and Eq.~(\ref{eq:zimpoka}) in conjunction with the neutrino beam
spectrum provided by the MiniBooNE collaboration~\cite{AguilarArevalo:2008yp}. For the LSND experiment, we used the analytical results for the relevant oscillation probabilities as given in Ref.~\cite{Ahluwalia:1996fy}. These expressions were designed to include the full spectral information of the neutrino beam and incorporated the detector energy cutoff of 52.8 MeV on the upper end, and 20 MeV on the lower end. \\

\textit{Results \textemdash~} 
In our analysis the MiniBooNE/LSND relevant mass-squared difference turns out to $\gtrsim  0.05\units{eV}^2$.
At the lower limit of this mass squared difference the limits on $\theta_{12}$ and $\theta_{23}$ are quite severe: $\theta_{12} \approx 0.6$ radians (which translates to $\theta_{13} \approx 0$ through Eq.~(\ref{eq:theta13}), and $\theta_{23}\approx 0$.
As one moves to a higher mass-squared difference a much larger, and perhaps more realistic, parameter space opens up.\\

For the extreme lower-limit case just mentioned one obtains a LSND-like~\textendash~but suppressed~\textendash~signal for MiniBooNE in the  $\nu_\mu\to\nu_e$  mode, while a significant number of $\nu_e$ events are spread over the whole energy range. These events are determined by the product of the $\nu_e$ cross section in the detector (with an appropriate compensation for efficiency), the second term on the right-hand side of the  Eq.~(\ref{eq:zimpoka}), and the $\nu_\mu$ flux at the source.  For the extreme case under consideration this integrated number  may be interpreted as a $\nu_\mu$ disappearance at the $20\%$ level. These details cannot, and should not, be interpreted as quantitative predictions, because the use of the SN1987a data requires a proper Bayesian analysis. They only define a strategy and provide qualitative insights. 
In the extreme lower-limit case outlined here the mixing matrix reduces to a $(2\times 2) \oplus (1\times 1)$ form and 
the SN1987a constraint is seen to restrict the allowed $(\sin^2(\theta),\Delta m^2)$ to the lower right-hand corner of the standard 
favoured region for LSND \cite{Aguilar:2001ty}.
\\

In this scenario, the MiniBooNE confirms a $\bar\nu_\mu\to\bar\nu_e$  LSND-like signal  and sees a suppressed
 $\nu_\mu\to\nu_e$  signal, with a significant number of $\nu_e$ events spread over the whole energy range.
\\

\textit{An alternate scenario~\textemdash}
As an alternate scenario we consider the possibility that $\mu^\pm$ decay dominates the decoherence process. In that event, the expectations for the MiniBooNE are revised to read (without an  intrinsic CP violation)
\begin{align}
  \chi\left(\bar{\nu}_\mu\to\bar{\nu}_e\right) = \chi\left(\nu_\mu\to\nu_e\right) = \exp[-\ell/\tau_\mu] \mathcal{P}\left(\nu_\mu\to\nu_e\right) +  \left(1- \exp[-\ell/\tau_\mu]\right) \sum_i \left(U^\ast_{\mu i} U_{\mu i} U_{e i} U^\ast_{ei} \right)
  \label{eq:zimpokj}
\end{align}
The oscillatory signal is suppressed while the constant contribution arising from the decoherence may be interpreted as a $\nu_e$ or $\bar\nu_e$ background. \\

\textit{Concluding remarks \textemdash}
The last decades have given increasingly stronger hints that neutrinos are not mass eigenstates. Each flavour is a different linear superposition of at least three different mass eigenstates. Goldman was the first to point out that this circumstance induces a source-dependent quantum entanglement. In the $\pi^\pm$ decay the $\nu_\mu$ and $\bar\nu_\mu$  are entangled with $\mu^+$ and $\mu^-$ respectively. But if their source is $\mu^\pm$ decay the entanglement is more intricate. For the reactor $\bar\nu_e$, e.g., the entanglement is with the parent neutron bound inside a nucleus. Clearly even for a given flavour of neutrino there is no unique entanglement. It depends on the source. The entanglement proposed by Goldman is not a matter of choice. It is a consequence of the conservation of the energy-momentum four vector and hence it must be incorporated in neutrino oscillation phenomenology. If one ignores the interaction of the entangled partners of neutrinos, and their specific properties, such as lifetimes, the new phenomenology returns the old to the lowest order in mass-squared differences. The point of departure for the present communication has been to explore the consequences if one refrains from ignoring such details. Choosing to work within the setting of the LSND experiment and the MiniBooNE, we brought forward the possibility that the two experiments may yield dramatically different results. Without invoking the source-dependence of the entanglement for neutrinos, and without considering the fate of the entangled partners as they evolve, much of the rich physics may be missed, and in fact, mis-interpreted. \\
 
We have shown that a distinct possibility exists in which a LSND-like experiment  sees no CP violation, whereas a MiniBooNE-like setup reproduces the LSND results for the $\bar\nu_\mu$ to $\bar\nu_e$ oscillations while seeing only a significantly suppressed signal for the $\nu_\mu$ to $\nu_e$ oscillations. Should such a possibility be realised experimentally, a new set of dedicated experiments would seem advised. At the same time it would become imperative that one looks at the existing experiments for their new potentialities. \\

\textit{Acknowledgements \textemdash} 
One of us (DVA) thanks Hywel White for bringing him back to the subject of neutrino oscillations, and for the ensuing discussions. 
We thank Cheng-Yang Lee for discussions in the early stages of this communication.
We also thank the Kaikoura Field Station of the University of Canterbury where much of this communication was written.
%\bibliography{Pion}
%Merlin.mbs v4.21 2009-07-09.
%Merlin.mbs v4.21 2009-07-09.
%

\end{document}